\documentclass[conference]{IEEEtran}
\IEEEoverridecommandlockouts

\usepackage{amsmath,amssymb}
\usepackage{graphicx}
\usepackage{booktabs}
\usepackage{multirow}
\usepackage{cite}
\usepackage{url}
\usepackage{placeins}
\usepackage{balance}

\graphicspath{{figures/icc/}}
\newcommand{\E}{\mathbb{E}}
\newcommand{\med}{\operatorname{med}}
\newcommand{\IQR}{\operatorname{IQR}}

\title{Progressive-Observation Physical-Layer Fingerprinting for Efficient Open-Set Ethernet Authentication}
\author{\IEEEauthorblockN{Chenming Zhang$^{1}$,~Jiaxiang Geng$^{1,2}$,~Aiqun Hu$^{3}$,~Azzedine Boukerche$^{4}$,~Peng Sun$^{1,4}$}
\IEEEauthorblockA{$^1$ Duke Kunshan University, China}
\IEEEauthorblockA{$^2$ The University of Hong Kong, China}
\IEEEauthorblockA{$^3$ Southeast University, China}
\IEEEauthorblockA{$^4$ University of Ottawa, Canada}}

\begin{document}
\AddToHookNext{shipout/foreground}{\put(0.5\paperwidth,-0.42in){\makebox[0pt]{\parbox{\textwidth}{\centering\footnotesize This work has been submitted to the IEEE for possible publication. Copyright may be transferred without notice, after which this version may no longer be accessible.}}}}
\maketitle

\begin{abstract}
Physical-layer fingerprinting (PLF) has emerged as a promising authentication approach for Internet of Things devices. However, most PLF systems acquire and process a fixed-length waveform for every authentication query, burdening resource-constrained edge nodes. To address this issue, we propose a progressive-observation PLF method that adapts the waveform length to each query. Specifically, a lightweight convolutional neural network (CNN) first rejects clearly unknown devices from a short observation. Spectral matching then accepts clearly registered devices from a slightly longer observation. Only ambiguous queries proceed to the same matching on the full-length waveform, reducing both data acquisition and computation. Because fingerprints drift across days, we also design an adaptive calibration algorithm to avoid retraining the CNN. The algorithm realigns the features and decision thresholds using a few current-day references from registered devices. We evaluate the method with 100BASE-TX Ethernet devices of two controller types on held-out days. Compared with fixed-length processing, it keeps the false rejection and false acceptance rates equal or lower for both controller types on both days. It also reduces the waveform data per query by 75.1\% and the CPU processing time on a Raspberry Pi 5 by 73.7\% on average. With the same monitoring bandwidth, an edge node can therefore serve about four times as many authentication queries.
\end{abstract}


\section{Introduction}

Many Internet of Things (IoT) deployments connect devices over wired Ethernet, where access control typically relies on IP and MAC addresses~\cite{zhongHPF2026}. These addresses are configured in software and can therefore be spoofed by an attacker. Physical-layer fingerprinting (PLF) addresses this weakness by authenticating a device through analog signal characteristics caused by manufacturing variations in its transmitter, clock, and analog front end~\cite{zhangTheory2025}. For Ethernet, PLF has progressed from matched-filter identification of 10/100-Mb/s cards~\cite{gerdesEthernet2012} to adaptive-filter~\cite{liAdaptive2021} and spectral~\cite{liuSpectrum2022} fingerprints. Recent 1000BASE-T studies add open-set recognition~\cite{zhongRisk2026} and interpretable fingerprint models~\cite{zhongHPF2026}.

Existing PLF research has advanced along two lines, namely authentication accuracy and processing efficiency. For accuracy, open-set methods reject unknown devices, for example by combining model- and data-driven features~\cite{xieOpenSet2021} or by verifying a predicted identity through Siamese comparison~\cite{caiOpenSet2025}. Other studies model time-varying fingerprint components~\cite{wangTime2024} and improve robustness to receiver changes~\cite{shenAgnostic2024,wangTriad2026} and limited acquisition diversity~\cite{chilletDatabase2024}. For efficiency, variable-length models accept different input sizes~\cite{shenLength2023}, lightweight features reduce storage and arithmetic~\cite{bothereauLightweight2026}, and early-exit networks stop model execution for easy samples~\cite{wenEarlyExit2023}.

However, most PLF systems still acquire and process a fixed-length waveform for each authentication query. Efficiency-oriented methods, including early-exit networks, reduce computation only after this waveform has been acquired and transferred. On resource-constrained edge nodes, this fixed per-query burden grows with the number of monitored devices and authentication requests. Moreover, fingerprints drift across acquisition days, so decision thresholds set on one day can become unreliable on later days. Retraining the model every day to track this drift is costly.

To reduce the per-query burden, a PLF system can acquire the waveform progressively and stop once the evidence is sufficient. In open-set authentication, rejecting an unknown device and accepting a registered device can require different amounts of evidence. Accordingly, a progressive PLF system must answer three questions for each query. First, it must determine when a short observation is sufficient to reject an unknown device. Second, it must determine when a slightly longer observation is sufficient to accept a registered device. Third, it must reliably decide the remaining ambiguous queries with a full-length observation.

\begin{figure}[t]
  \centering
  \includegraphics{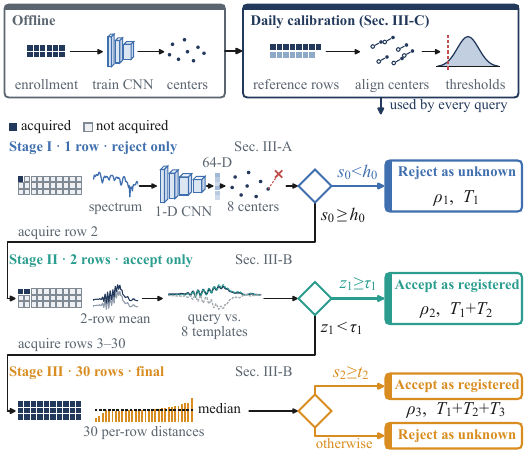}
  \caption{Method overview: offline phase, daily calibration, and three-stage progressive cascade. $\rho_j$ and $T_j$ are the termination probability and processing time of stage $j$ in~\eqref{eq:datareq}. The spectra are measured, and the other glyphs are schematic.}
  \label{fig:pipeline}
\end{figure}

We propose a progressive-observation PLF method that combines a three-stage cascade with an adaptive calibration algorithm (Fig.~\ref{fig:pipeline}). The cascade starts from one waveform row and acquires more rows only for queries that the current stage leaves undecided. Stage~I answers the first question with a lightweight one-row convolutional neural network (CNN) that only rejects clearly unknown devices. Its convolutions respond to the shape of local frequency bands, whereas spectral matching compares individual frequency bins. Stage~II answers the second question by adding one row and applying spectral matching that only accepts clearly registered devices. Stage~III answers the third question by applying the same matching to the full-length observation of the remaining queries. These roles restrict each short observation to the decision it supports reliably. Each early decision also saves acquiring, transferring, and processing the remaining rows.

Furthermore, the adaptive calibration algorithm keeps the decision thresholds valid across days while reusing the trained CNN. It realigns the CNN embeddings and spectral features and updates the thresholds using a few current-day references from registered devices.

The main contributions of this paper are summarized as follows.
\begin{itemize}
\item We propose a three-stage progressive-observation PLF method with stage-specific decision roles, so most queries terminate after one or two waveform rows.
\item We design an adaptive calibration algorithm that keeps the decision thresholds valid across days without retraining the CNN.
\item We evaluate the method with 100BASE-TX Ethernet devices of two controller types on held-out days. Compared with fixed-length processing, it keeps the false rejection and false acceptance rates equal or lower while reducing waveform data by 75.1\% and CPU processing time by 73.7\% on average.
\end{itemize}

In the rest of this paper, Section~\ref{sec:formulation} defines the observation depth and processing cost and formulates the threshold design problem. Section~\ref{sec:method} presents the three-stage cascade and the adaptive calibration algorithm. Section~\ref{sec:setup} describes the measurement set-up and evaluation protocol, Section~\ref{sec:results} reports authentication performance and resource savings, and Section~\ref{sec:conclusion} concludes the paper.

\section{Problem Formulation}
\label{sec:formulation}

Consider $K$ registered devices with identities $\mathcal K=\{1,\ldots,K\}$, and let label 0 denote an unknown device. Each authentication query can acquire up to $M$ waveform rows of $L$ samples and returns either label 0 or an identity in $\mathcal K$. A progressive policy processes the query in $J$ stages with nested observations of $n_1<\cdots<n_J=M$ rows. Its thresholds $\theta$ determine the probability $\rho_j$ that a query terminates at stage $j$. Because later stages reuse the earlier rows,
\begin{equation}
 \begin{aligned}
 \E[N]&=\textstyle\sum_{j=1}^{J}\rho_jn_j,\qquad S_{\rm data}=1-\E[N]/M,\\
 C(\theta)&=\textstyle\sum_{j=1}^{J}T_j\sum_{i=j}^{J}\rho_i,
 \end{aligned}
 \label{eq:datareq}
\end{equation}
where $T_j$ is the processing time of stage $j$ and $C$ is the route-weighted processing time per query. At a fixed row encoding and query rate, the monitoring-waveform data volume per query and the link data rate both scale with $\E[N]$.

Let $\Delta_m(\theta)$ denote the degradation of open-set metric $m\in\mathcal M$ relative to fixed-length processing, which acquires all $M$ rows for every query. Each $\Delta_m$ adds a bootstrap upper margin to the mean degradation over CNN seeds. The design problem is
\begin{equation}
 \min_{\theta\in\Theta}\ C(\theta)\quad\text{s.t.}\quad \Delta_m(\theta)\le\epsilon,\ \ \forall m\in\mathcal M,
 \label{eq:objective}
\end{equation}
where $\epsilon$ is the tolerated degradation and $\Theta$ is a finite grid of stage thresholds. Every candidate restricts each early stage to one decision type (Section~\ref{sec:decision}). We enumerate $\Theta$ on development data and keep a low-cost feasible candidate. Section~\ref{sec:method} instantiates this policy as a three-stage cascade with $(n_1,n_2,n_3)=(1,2,M)$. One row is the smallest unit that a query can acquire. Two rows are the fewest that allow spectral averaging, which reduces row-level variation. The $M$ rows form the full-length observation that fixed-length processing uses.

\section{Progressive-Observation PLF and Cross-Day Calibration}
\label{sec:method}

Figure~\ref{fig:pipeline} summarizes the resulting cascade, which comprises a one-row CNN stage, a two-row spectral stage, and an $M$-row spectral stage. The calibration in Section~\ref{sec:calibration} adapts the neural and spectral references to each acquisition day.

\subsection{One-Row CNN Representation}

From each row, $W$ evenly spaced Hann-windowed segments of $L_{\rm s}$ samples are converted to log spectra $\mathbf x_q$ and passed through a compact 1-D CNN $g_\phi(\cdot)$. The normalized row embedding is
\begin{equation}
 \begin{aligned}
 \mathbf e&=\operatorname{norm}\!\left(\frac{1}{W}\sum_{q=1}^{W}
 \operatorname{norm}\big(g_\phi(\mathbf x_q)\big)\right),\\
 \mathcal L&=\mathcal L_{\rm supcon}+\gamma\mathcal L_{\rm CE}.
 \end{aligned}
 \label{eq:cnn}
\end{equation}
Here $\mathcal L_{\rm supcon}$ is the supervised contrastive objective~\cite{khoslaSupCon2020}, $\mathcal L_{\rm CE}$ is cross-entropy, and $\gamma$ is its weight. The CNN is trained on enrollment captures of registered devices with gain, noise, and spectral-tilt augmentation.

For identity $k$, $\mathbf c_k$ is the normalized mean enrollment embedding, $\eta_k=\overline{(1-\mathbf e^\mathsf T\mathbf c_k)}$ is its mean within-identity distance, and $\sigma_k=\max\{\eta_k,\med_\ell\eta_\ell\}+10^{-6}$ is the stabilized scale. Stage~I computes
\begin{equation}
 d_{0,k}=\frac{1-\mathbf e^\mathsf T\widetilde{\mathbf c}_k}{\sigma_k},\qquad
 s_0=-\min_k d_{0,k},
 \label{eq:cnnscore}
\end{equation}
where $\widetilde{\mathbf c}_k$ is the day-aligned center defined in Section~\ref{sec:calibration}. When $s_0$ falls below the Stage~I rejection threshold $h_0$, the row is far from every registered identity and the query is rejected as unknown. Otherwise, the first row is retained and acquisition continues.

\subsection{Two-Row and Full-Length Spectral Evidence}\label{sec:spectral}

Let $\mathbf P_r$ be the \emph{linear} power spectrum of row $r$. The spectral representation comprises a feature location $\boldsymbol\mu$, scale $\boldsymbol\alpha$, diagonal variance $\mathbf v$, and registered-device centers $\mathbf q_k$. Its distance to identity $k$ is
\begin{equation}
 D(\mathbf P,k)=\frac{1}{F}\sum_{f=1}^{F}
 \frac{\left([\log_{10}\mathbf P-\boldsymbol\mu]_f/\alpha_f-q_{k,f}\right)^2}{v_f}.
 \label{eq:distance}
\end{equation}
Stage~II averages the first two power spectra before taking the logarithm,
\begin{equation}
 s_1=-\min_kD\!\left(\tfrac{\mathbf P_1+\mathbf P_2}{2},k\right),
 \quad p_1=\arg\min_kD_u\!\left(\tfrac{\mathbf P_1+\mathbf P_2}{2},k\right),
 \label{eq:stage2}
\end{equation}
where $D_u$ is~\eqref{eq:distance} without the variance weights $\mathbf v$. The weighted distance sets the acceptance score, while $D_u$ only assigns the nearest identity. If the calibrated score $z_1$ (Section~\ref{sec:calibration}) reaches the Stage~II acceptance threshold $\tau_1$, the query is accepted as $p_1$, and otherwise it is extended to $M$ rows.

Stage~III aggregates per-row distances rather than averaging the spectra:
\begin{equation}
 \begin{aligned}
 \bar D_k&=\med_{r=1}^{M}D(\mathbf P_r,k),\\
 p_2&=\arg\min_k\bar D_k,\qquad s_2=-\min_k\bar D_k.
 \end{aligned}
 \label{eq:stage3}
\end{equation}
The median reduces the influence of atypical rows. Stage~III accepts the predicted identity $p_2$ when $s_2$ reaches its identity-specific threshold $t_{2,p_2}$ and rejects the query as unknown otherwise.

Stage~I uses a CNN because its evidence differs from that of~\eqref{eq:distance}. The nonlinear and impulse-response model (NAIM)~\cite{zhangNAIM2026} describes the waveforms of AX88772 and RTL8152B adapters by steady-state level deviations and a transition response built from damped sinusoids. Each damped sinusoid is concentrated near its center frequency and thus changes a band of adjacent bins. The distance~\eqref{eq:distance} weights each bin separately, whereas the convolutions over $\mathbf x_q$ respond to the shape of such bands.

\subsection{Adaptive Cross-Day Calibration}
\label{sec:calibration}

Each acquisition day provides a prespecified reference block of registered-device rows, kept separate from the evaluation queries. The reference embeddings estimate a shared shift $\boldsymbol\delta$ between the enrollment and current-day feature spaces, yielding $\widetilde{\mathbf c}_k=\operatorname{norm}(\mathbf c_k+\boldsymbol\delta)$ in~\eqref{eq:cnnscore}. The same block updates the identity centers $\mathbf q_k$ and the variance $\mathbf v$ used in~\eqref{eq:distance}, while $\boldsymbol\mu$ and $\boldsymbol\alpha$ keep their enrollment values. Thus, the trained CNN and the spectral distance rule remain unchanged, and only their reference representations are aligned with the current acquisition day.

Each score is standardized against the median and interquartile range (IQR) of the registered-reference scores $G_j$ at stage $j$~\cite{rousseeuwRobust2011},
\begin{equation}
 z_j=\frac{s_j-\med(G_j)}{\max\{\IQR(G_j),10^{-6}\}}.
 \label{eq:z}
\end{equation}
The Stage~I rejection threshold is $h_0=\beta_0-\lambda_0\omega_0$, where $\beta_0$ and $\omega_0$ are a robust location and a robust spread of $G_0$ after an enrollment-day offset, and $\lambda_0$ sets the rejection margin.

The identity-specific Stage~III threshold also accounts for systematic differences among registered identities:
\begin{equation}
 t_{2,k}=\med(G_{2,k})+\big[t_{2,\rm pool}-\med(G_2)\big].
 \label{eq:cardloc}
\end{equation}
Here, the pooled Stage~III threshold $t_{2,\rm pool}$ is shared by all identities, and~\eqref{eq:cardloc} shifts it according to the current-day median score of the predicted identity. The grid $\Theta$ in~\eqref{eq:objective} spans $\lambda_0$, $\tau_1$, and two candidate rules for $t_{2,\rm pool}$. The first minimizes the sum of the false rejection rate (FRR) and false acceptance rate (FAR) estimated on resampled reference rows. The second sets $t_{2,\rm pool}=\med(G_2)-\lambda_2\IQR(G_2)$ with margin $\lambda_2$.

\subsection{Stage-Specific Decision Roles}
\label{sec:decision}

The terminal decision is
\begin{equation}
 (\hat o,\hat y)=
 \begin{cases}
 (\mathrm{unknown},0), & s_0<h_0,\\
 (\mathrm{known},p_1), & s_0\ge h_0,\ z_1\ge\tau_1,\\
 (\mathrm{known},p_2), & s_0\ge h_0,\ z_1<\tau_1,\ s_2\ge t_{2,p_2},\\
 (\mathrm{unknown},0), & \text{otherwise.}
 \end{cases}
 \label{eq:routing}
\end{equation}
The first case rejects the query after one row, and the second case accepts it as $p_1$ after two rows. The remaining queries reach Stage~III, which uses all $M$ rows to accept them as $p_2$ (third case) or reject them (fourth case). These cases define the termination probabilities $\rho_j$ in~\eqref{eq:datareq} and hence the mean observation depth $\E[N]$.

Stage~I therefore only rejects, Stage~II only accepts, and Stage~III can make either decision. This assignment reflects the different evidence that the two decision types require. A rejection requires only that the query lie far from every registered identity. An acceptance also requires the query to match one identity, so Stage~II makes it from the averaged two-row spectrum and the calibrated score $z_1$.

The assignment also separates the error types of the two early stages. Every Stage~I error is a false rejection of a registered device. Every Stage~II error is either a false acceptance of an unknown device or a wrong-identity acceptance of a registered device. Thus, $h_0$ governs the rejection errors of Stage~I, and $\tau_1$ governs the acceptance errors of Stage~II. The constraint in~\eqref{eq:objective} bounds the resulting degradation of every open-set metric on development data. Section~\ref{sec:roles} compares this restricted assignment with an unrestricted variant in which Stages~I and~II can each accept or reject.

\section{Experimental Methodology}
\label{sec:setup}

\begin{figure}[t]
  \centering
  \includegraphics{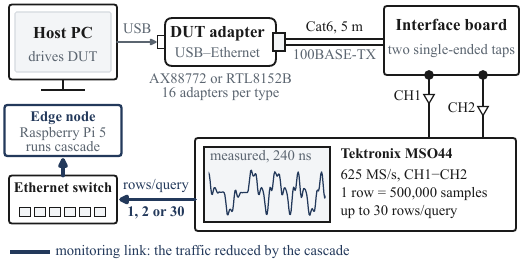}
  \caption{Testbed. Acquisition chain and monitoring link from the oscilloscope to the edge node. The displayed waveform is measured.}
  \label{fig:testbed}
\end{figure}

Figure~\ref{fig:testbed} shows the testbed, in which a host PC drives the 100BASE-TX USB-to-Ethernet adapter under test (DUT) through USB. A 5-m Cat6 UTP cable forms the wired channel between the DUT and an interface board that exposes two single-ended taps. A Tektronix MSO44 samples CH1 and CH2 at 625 MS/s and forms the CH1$-$CH2 differential waveform. A wired Ethernet switch returns this waveform to the edge node, a Raspberry Pi 5 that runs the cascade. As in~\cite{zhangNAIM2026}, each stored capture contains $M=30$ differential-waveform rows, each with $L=500{,}000$ samples. For both controller types, Stage~I uses one row acquired after the link stabilizes. The cascade acquires rows progressively, so each query returns only 1, 2, or 30 rows over the monitoring link. The evaluation replays these rows from the stored 30-row captures.

\begin{figure}[t]
  \centering
  \includegraphics{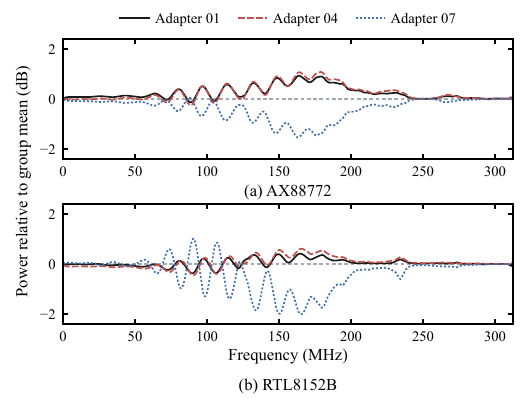}
  \caption{Spectra of three registered adapters per controller type, shown as the deviation from the mean spectrum of all eight registered adapters of that type. Curves are smoothed over 21 frequency bins for display.}
  \label{fig:signal}
\end{figure}

We study USB-to-Fast-Ethernet adapters based on the AX88772 and RTL8152B controllers. For each controller type, eight physical adapters are registered and eight are treated as unknown. Figure~\ref{fig:signal} shows the spectra of three registered adapters per controller type, relative to the mean spectrum of that type. Two consecutive captures (A and B) of each adapter yield $Q=32$ query captures per controller type and acquisition day. Captures from the first two days train the CNN with $\gamma=0.3$ and establish the enrollment spectral model. The CNN takes $W=30$ segments of $L_{\rm s}=8{,}192$ samples per row, outputs 64-D embeddings, and has 52,272 and 81,768 trainable parameters for AX88772 and RTL8152B, respectively. The complete method, including all thresholds and cross-stage score mappings, is developed on the third day, whose captures form the development data. It is then applied to held-out query captures from the fourth and fifth days without further model or parameter selection.

We set $\epsilon=0.3$ percentage points and run the search in~\eqref{eq:objective} once per controller type. It selects $(\lambda_0,\tau_1)=(6,-0.5)$ and the error-minimizing rule for AX88772, and $(4,-1)$ and the margin rule with $\lambda_2=1.5$ for RTL8152B. For each registered identity, the reference block of Section~\ref{sec:calibration} contains eight rows that align the neural and spectral references. Seven further rows provide the registered-reference scores $G_j$, giving $R=120$ shared rows per day.

The fixed-length processing baseline shares the current-day calibration, 30-row spectral representation, identity prediction, and identity-specific threshold with Stage~III. This design isolates the effects of the two early decisions and of progressive termination. Each evaluation is repeated with different CNN seeds, trials, and A/B reference--query orders on the same captures, and all reported rates are computed over these repetitions.

Authentication metrics are FRR, FAR, wrong-identity acceptance (WIA), the area under the receiver operating characteristic curve (AUROC), and the open-set classification-rate curve (OSCR)~\cite{dhamijaOSCR2018}. FRR is the fraction of queries from registered devices that are rejected, and FAR is the fraction of queries from unknown devices that are accepted. WIA is the fraction of queries from registered devices that are accepted under a wrong identity. OSCR relates correct registered-device identification to unknown-device false acceptance as the decision threshold varies. For the cascade, AUROC and OSCR require a common ranking of decisions from different stages. A positive-slope affine mapping $e_j=a_jz_j+b_j$ is fitted for each controller type by logistic regression on development-day registered-versus-unknown labels. The common ranking score is $e_0$ for a Stage~I rejection, $\max\{e_1,e_2(t_{2,\rm pool})\}$ for a Stage~II acceptance, and $e_2$ for a deferred query. For Stage~II, $e_2(t_{2,\rm pool})$ is a fixed offset derived from the pooled Stage~III threshold.

Efficiency metrics are the monitoring-waveform rows and CPU processing time per query. CPU processing time is measured on a Raspberry Pi 5 (8 GB RAM) from the acquired rows to the terminal decision. Each stage is also timed separately as the median of 18 repetitions. Terminal-path times are weighted by the route fractions observed on the fourth and fifth days, while the baseline executes the same 30-row Stage~III implementation. Monitoring-waveform data volume is calculated from the terminal row counts through~\eqref{eq:datareq}. The daily data reduction $1-(R+Q\E[N])/(R+MQ)$ also counts the $R$ shared reference rows.

\section{Results}
\label{sec:results}
\subsection{Open-Set Authentication Performance}

Table~\ref{tab:quality} compares the cascade with uniform 30-row processing on the four controller--day combinations. OSCR equals AUROC in every row, and bold marks error rates lower than those of the baseline. The cascade improves one of the two open-set error rates in three of the four combinations and leaves both unchanged in the fourth. FAR decreases on both AX88772 days, while FRR decreases on the fourth RTL8152B day. FRR and FAR remain equal or lower in every combination, and WIA is zero throughout.

The threshold-independent metrics, AUROC and OSCR, remain nearly identical for the two methods. Both methods achieve 100\% AUROC and OSCR on the AX88772 data, while their differences on the two RTL8152B days are 0.052 and 0.069 percentage points. The proposed method therefore improves the selected operating point while leaving the open-set ranking nearly unchanged.

\begin{table}[t]
\centering
\caption{Authentication performance on days 4 and 5.}
\label{tab:quality}
\footnotesize
\setlength{\tabcolsep}{4pt}
\renewcommand{\arraystretch}{1.02}
\begin{tabular}{llccc}
\toprule
Type/day & Method & AUROC (\%) & FRR (\%) & FAR (\%)\\
\midrule
AX/D4  & Uniform 30-row & 100.00 & 0.00 & 2.30\\
       & Cascade (ours) & 100.00 & 0.00 & \textbf{1.97}\\
AX/D5  & Uniform 30-row & 100.00 & 0.00 & 0.66\\
       & Cascade (ours) & 100.00 & 0.00 & \textbf{0.00}\\
RTL/D4 & Uniform 30-row & 100.00 & 0.63 & 0.00\\
       & Cascade (ours) & 99.95  & \textbf{0.00} & 0.00\\
RTL/D5 & Uniform 30-row & 100.00 & 4.58 & 0.00\\
       & Cascade (ours) & 99.93  & 4.58 & 0.00\\
\bottomrule
\end{tabular}
\end{table}

\subsection{Why Stage-Specific Decision Roles Matter}
\label{sec:roles}

Table~\ref{tab:authority} separates the contribution of the two early decisions over the four controller--day combinations. OSCR again equals AUROC, and WIA is zero for all variants. Using only the Stage~I rejection lowers mean FAR from 0.74\% to 0.49\%, while using only the Stage~II acceptance lowers mean FRR from 1.30\% to 1.15\%. Combining them retains both improvements and reduces the mean observation depth from 30 to 7.47 rows. Therefore, early termination alone does not explain the gain in authentication performance. The gain comes from assigning each short observation only the decision that it supports reliably. Stage~I, for example, rejects some unknown queries that 30-row spectral matching accepts, consistent with their different evidence (Section~\ref{sec:spectral}).

\begin{table}[t]
\centering
\caption{Ablation of stage-specific decision roles on days 4 and 5.}
\label{tab:authority}
\footnotesize
\setlength{\tabcolsep}{4pt}
\renewcommand{\arraystretch}{1.05}
\begin{tabular}{lcccc}
\toprule
 & Worst & \multicolumn{3}{c}{Mean}\\
\cmidrule(l){3-5}
Variant & AUROC (\%) & FRR (\%) & FAR (\%) & Rows\\
\midrule
Uniform 30-row       & 100.00 & 1.30 & 0.74 & 30.00\\
Stage~I reject only  & 99.90  & 1.30 & 0.49 & 18.74\\
Stage~II accept only & 100.00 & 1.15 & 0.74 & 18.73\\
\textbf{Restricted (ours)} & \textbf{99.93} & \textbf{1.15} & \textbf{0.49} & \textbf{7.47}\\
Unrestricted         & 91.51  & 2.94 & 2.77 & 2.82\\
\bottomrule
\end{tabular}
\end{table}

The unrestricted variant also selects its Stage~I acceptance and Stage~II rejection thresholds on the third acquisition day. It loses at most 0.3 percentage points of mean development performance relative to uniform 30-row processing. Although this variant reduces the mean observation depth to 2.82 rows, its worst AUROC/OSCR on the subsequent days falls to 91.51\%, and its mean FRR/FAR rise to 2.94\%/2.77\%. Short observations can therefore appear efficient on development data yet make unreliable decisions after the acquisition conditions change. Restricting their decision roles keeps the early decisions reliable across days.

\begin{figure}[t]
  \centering
  \includegraphics{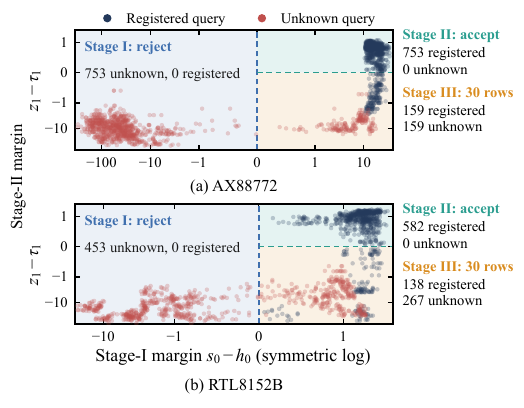}
  \caption{Decision regions of Stages~I and~II on the fifth acquisition day for (a) AX88772 and (b) RTL8152B. Points are repeated evaluations of the same captures, and counts give the evaluations ending in each region.}
  \label{fig:decision}
\end{figure}

Figure~\ref{fig:decision} shows how the two early thresholds divide the queries. Registered queries concentrate above the Stage~II acceptance threshold, unknown queries concentrate below the Stage~I rejection threshold, and conflicting or insufficient evidence is deferred to Stage~III. Pooled over the fourth and fifth acquisition days, Stage~I rejects 74--83\% of unknown queries and Stage~II accepts 76--86\% of registered queries for the two controller types. Every Stage~I rejection is an unknown query, and every Stage~II acceptance is a registered query.

\subsection{Resource Savings from Progressive Observation}
\begin{figure*}[t]
  \centering
  \includegraphics{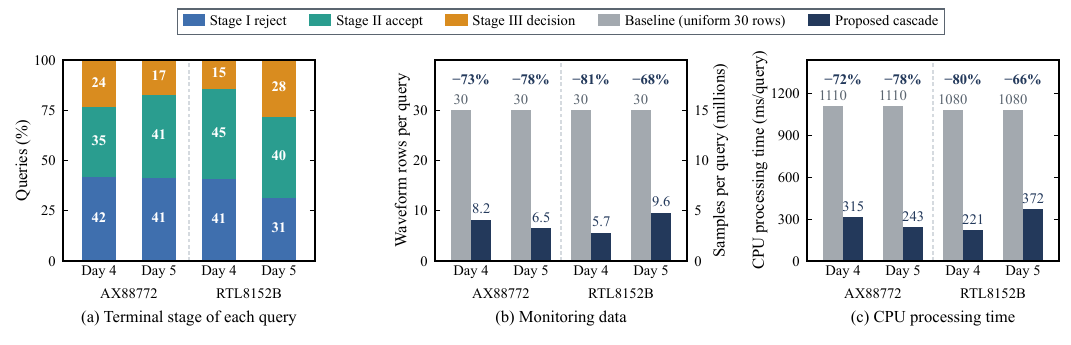}
  \caption{Routing and resource results on the fourth and fifth acquisition days. (a) Terminal-stage shares. (b) Waveform rows and samples per query. (c) Route-weighted CPU processing time per query. Bold labels give the reduction relative to the uniform 30-row baseline.}
  \label{fig:results}
\end{figure*}

Most queries terminate at the first two stages, which lowers the cost of each query (Fig.~\ref{fig:results}). Across the four controller--day combinations, Stage~I terminates 31.46--41.67\% of queries and Stage~II another 34.70--44.53\% [Fig.~\ref{fig:results}(a)]. The mean observation depth falls from 30 to 5.66--9.56 rows, reducing monitoring-waveform data volume per query by 75.1\% on average [Fig.~\ref{fig:results}(b)]. With the shared reference block included, the daily data reduction remains 60.56--72.12\%. CPU processing time falls from 1080--1110~ms to 221--372~ms per query, a reduction of 73.7\% on average [Fig.~\ref{fig:results}(c)]. The component times are 22.9--31.8~ms for Stage~I and 1078.9--1117.8~ms for Stage~III, so the queries that reach Stage~III dominate the CPU processing time.

\section{Conclusion}
\label{sec:conclusion}

This paper proposed progressive-observation PLF for open-set 100BASE-TX authentication. Its three stages increase observation depth with query difficulty and make only decisions supported by their evidence. Current-day registered-device references update the neural and spectral representations and decision thresholds, so the trained CNN is reused across days. On held-out days, the method maintains or improves authentication across two Ethernet controller types while reducing monitoring-waveform data volume by 75.1\% and measured CPU processing time by 73.7\% on average. Future work will implement live progressive acquisition, measure its latency and energy, and test cross-scenario wireless radio-frequency fingerprinting with edge inference and cloud-coordinated adaptation.

\bibliographystyle{IEEEtran}
\bibliography{references}

\end{document}